\documentclass{article}
\usepackage{dcase2019,amsmath,graphicx,url}
\usepackage{times,booktabs, tabularx, url, amsfonts}
\usepackage{setspace,subcaption,etoolbox}

\title{Crowdsourcing a Dataset of Audio Captions}
\name{Samuel Lipping, Konstantinos Drossos, and Tuomas Virtanen}
\address{Audio Research Group, Tampere University, Tampere, Finland\\
        \{firstname.lastname\}@tuni.fi}

\begin{document}
\ninept
\maketitle
\begin{sloppy}
\begin{abstract}
Audio captioning is a novel field of multi-modal translation and it is the task of creating a textual description of the content of an audio signal (e.g. ``people talking in a big room''). The creation of a dataset for this task requires a considerable amount of work, rendering the crowdsourcing a very attractive option. In this paper we present a three steps based framework for crowdsourcing an audio captioning dataset, based on concepts and practises followed for the creation of widely used image captioning and machine translations datasets. During the first step initial captions are gathered. A grammatically corrected and/or rephrased version of each initial caption is obtained in second step. Finally, the initial and edited captions are rated, keeping the top ones for the produced dataset. We objectively evaluate the impact of our framework during the process of creating an audio captioning dataset, in terms of diversity and amount of typographical errors in the obtained captions. The obtained results show that the resulting dataset has less typographical errors than the initial captions, and on average each sound in the produced dataset has captions with a Jaccard similarity of 0.24, roughly equivalent to two ten-word captions having in common four words with the same root, indicating that the captions are dissimilar while they still contain some of the same information. 
\end{abstract}
%
%
\section{Introduction}\label{sec:intro}
Multimodal datasets usually have a set of data in one modality and paired set of data in another modality, creating an association of two different forms of media. These datasets differ from a typical classification or regression dataset in the sense that the two modalities convey the same content, but in different form. One example is the captioning task, where the datasets include a form of media (e.g. image) and then a textual description of the perceived content of the media. Two examples of captioning are image~\cite{karpathy:2017:tpami,chen15, young14} and audio captioning~\cite{drossos17}, where a textual description, i.e. a caption, is generated given an image or an audio file, respectively. Captioning tasks largely use deep learning methods~\cite{he17, amaresh19} where models are trained and evaluated on datasets, rendering the dataset as an important factor for the quality of the developed methods. Therefore, a good captioning dataset should have captions that are able to represent the differences on the perceived content (i.e. diverse captions). Also, it should have multiple captions per sample in order to represent the different ways of writing the same information (i.e. rephrasing) and allowing for a better assessment of the performance of the captioning method~\cite{yi14}.

Different datasets exist for image captioning~\cite{chen15, young14, hodosh13}, consisting of images and multiple captions per image. Most (if not all) of image captioning datasets are created by employing crowdsourcing and annotators that are located in an English speaking area (e.g. U.S.A., Australia, U.K., etc). Crowdsourcing provides several benefits, such as having no restrictions on location and the possibility of simultaneous annotation, and using a ready crowdsourcing platform provides the additional benefit of having an established base of users, i.e. potential annotators~\cite{drossos14}. One example of an image captioning dataset is the Flickr 8K dataset, which consists of 8092 images with five captions each and the captions were obtained by crowdsourcing~\cite{hodosh13}. The dataset images were hand-selected and they depict actions and events to encourage full sentence captions. The annotators were pre-screened (by answering questions regarding grammar and image captioning), and were required to be located in the US and to have an approval rate of 95\% on previous tasks on the crowdsourcing platform~\cite{rashtchian10}. Another example of a crowsourced image captioning dataset is the Microsoft COCO Caption dataset, which consists of five captions for over 200 000 images~\cite{yi14}. Annotators captioned the images one at a time and were told to write captions that contain at least eight words~\cite{chen15}. The restriction for the eight words encourages describing the image thoroughly and benefits the diversity of captions. 

Audio captioning is a novel recent area of research, introduced in~\cite{drossos17}. The first dataset used for audio captioning in~\cite{drossos17} is proprietary and consists of textual descriptions of general audio~\cite{pse:2015:pse}. Recently two other audio captioning datasets have been created, which are not proprietary; the Audio Caption and the AudioCaps datasets. The Audio Caption~\cite{wu19} dataset was partially released with 3710 video clips and their audio tracks, annotated with Mandarin Chinese and English captions. Each video had a duration of about 10 seconds and was annotated with three distinct captions. The video clips were annotated by Chinese university students. Annotators were instructed to focus on the sound of the video clips. The Chinese captions were then translated to English with Baidu Translation~\cite{wu19}. The AudioCaps~\cite{audiocaps} dataset was created by crowdsourcing. The audio material consists of 46 000 audio files from 527 audio event categories from the AudioSet~\cite{gemmeke17} dataset, and each audio file in AudioCaps has been annotated with one caption. The annotators were pre-screened by discerning those participants who consistently violated the given instructions, such as transcribing speech in the audio or describing the visual stimulus instead of the audio. Additionally, the annotators were required to be located in an English-speaking area, to have an approval rate of 95\% on previous tasks, and to have at least 1000 approved submissions on the crowdsourcing platform. Audio clips were selected such that the distribution of their classes was balanced, while ignoring classes that require visuals to be identified (for example the category ``inside small room'').

There are some considerations regarding the above mentioned datasets. Sound ambiguity is a known and well exploited property (for example in folley sounds) and by providing visual stimuli or word indications there is a high chance of reducing ambiguity and hampering the diversity of the captions. In both datasets, the diversity is hampered by the lack of a minimum amount of words in a caption, while in the case of AudioCaps the diversity is further hampered by removing informative sounds about spatial relationships (e.g. ``inside small room''), and giving the labels from AudioSet to the annotators to guide the caption. Additionally, Audio Caption is created by annotators who are not located in an English speaking area, increasing the chance for flaws in the acquired English captions. Finally, in the AudioCaps the assessment of the audio captioning methods is hindered by having only one caption per sound. 

In this paper we set out to provide a framework for annotating large amounts of audio files with multiple captions, which is structured in three steps. We have used this framework in preparing a new and freely available audio captioning dataset, which will be released during Autumn 2019. We employ the Amazon Mechanical Turk (AMT) as our crowdsourcing platform which has been used in numerous previous studies~\cite{chen15, rashtchian10, audiocaps, marge10, sorokin08}. The rest of the paper is organized as follows. In Section~\ref{sec:propmethod} we describe the proposed framework and in Section~\ref{sec:evaluation} we describe how we evaluate the effectiveness of the structure of our framework. The results of the evaluation are presented in Section~\ref{sec:results}. The paper is concluded in Section~\ref{sec:conclusions}.
%
%
\section{Proposed Framework}
\label{sec:propmethod}
Our proposed framework consists of three serially executed steps, inspired by practices followed in the creation of image captioning and machine translation datasets~\cite{hodosh13,yi14,zaidan11}, and which is implemented on an online crowdsourcing platform. We employ as potential annotators the registered users of the platform (the total amount of registered users is not fixed and depends on the platform) who are located in an English speaking area and have at least 3000 and 95\% approved (i.e. total and approval rate) submissions on the platform. From the potential annotators, annotators are selected for the steps on a first-come-first-served basis. Again, the number of potential annotators is not known and depends on the platform used. The assignment of audio files and captions to annotators will be explained in each step. The framework employs a set of $N_{\text{a}}$ audio files and produces a set of $N_{\text{c}}$ captions per file. In the first step we gather $N_{\text{c}}$ initial audio captions per file and in the second step the initial $N_{\text{c}}$ captions are edited, resulting to a second set of $N_{\text{c}}$ edited captions per audio file. In the third step we select the best $N_{\text{c}}$ captions per file between the initial and the edited ones. After each step, we screen the answers of the annotators. We permanently exclude annotators that consistently do not follow the given instructions, and we re-assign work that is deemed unacceptable in the screening process to other annotators. The workflow for our proposed framework is visualized in Figure~\ref{fig:amt-task-flow}.

In more detail, in the first step we solicit annotators for producing captions for all $N_{\text{a}}$ audio files. An annotator is presented with an audio file that is randomly selected from the audio files that have not yet been annotated and asked to write one caption, i.e. a sentence describing the perceived contents of the audio file without assuming any information not present in the audio. No other information is provided to the annotators to aid in describing the audio stimulus (i.e. no access to the name of the audio file, to any tags, or to any visual information is given). This way the caption contains only the perceived information from the audio stimulus and is not based on any prior knowledge about the audio file. To encourage providing descriptive captions with adequate information, we set a minimum caption length of eight words. Additional instructions for annotators in the first task include not to use non-descriptive phrases, such as ``There is'', ``I hear'', ``sound of'', and ``sounds like'' in the caption to reach the limit of eight words. We present each of our audio files to $N_{\text{c}}$ distinct annotators. From the first step we acquire a set of $N_{\text{c}}$ initial captions for each of the audio files employed.
\begin{figure}[!t]
    \centering
    \includegraphics[width=.85\columnwidth]{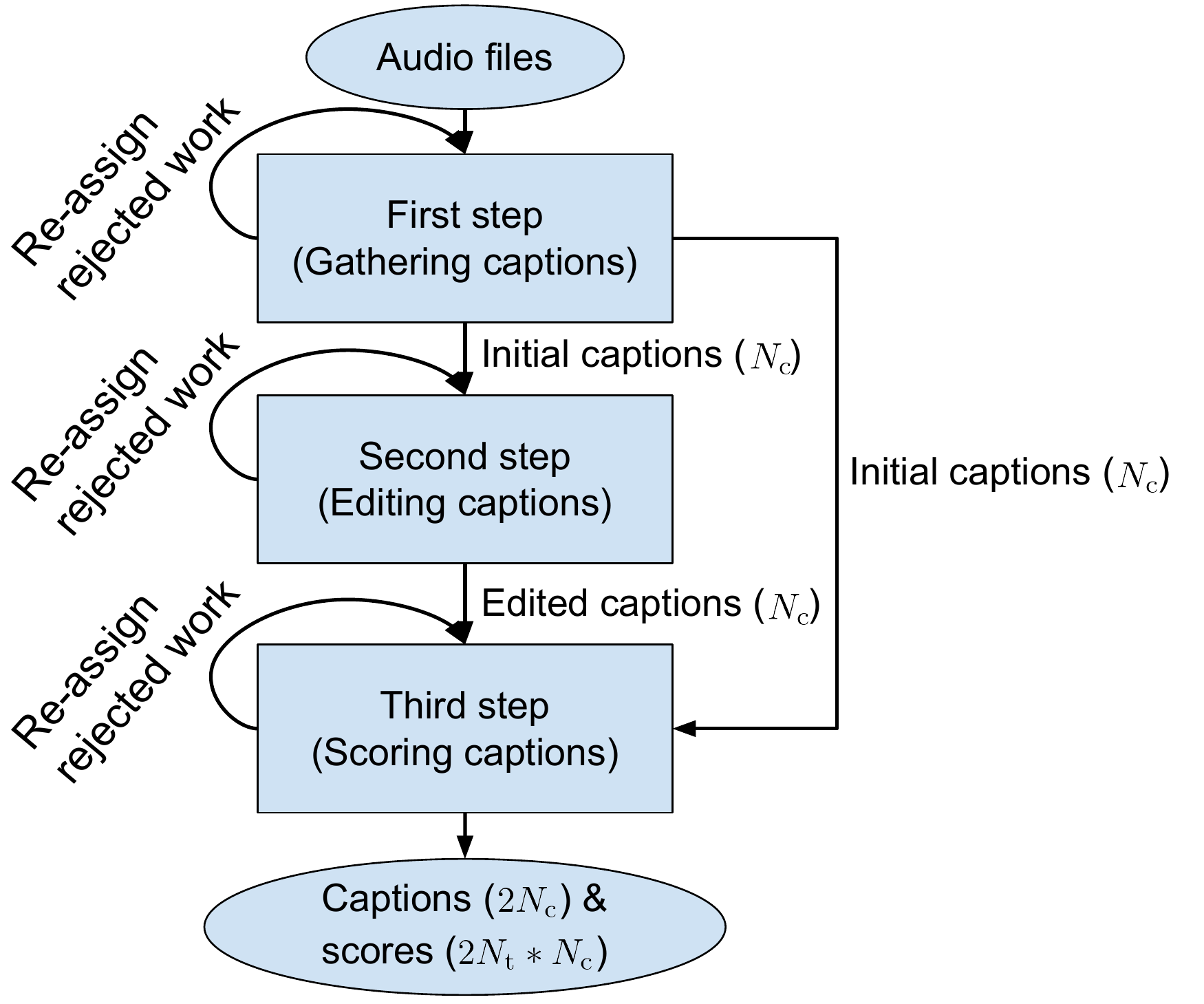}
    \caption{Graph of the task flow of our framework.}
    \label{fig:amt-task-flow}
\end{figure}

Some of the initial captions might include grammatical and/or typographical errors (e.g. ``\emph{An} car is driving...''), awkward sentence structures (e.g. ``An bird swallow-squelches to itself in an small branch''), or similar problems that are easier for humans to detect than for an algorithm. For that reason we introduce a second step for crowdsourcing the correction of any errors in the initial captions, where each initial caption is edited once. In this step an annotator is presented with a random caption that has not yet been edited from the initial captions (i.e. the annotator does not have access to any other information but only the caption). The annotator is instructed to read the given caption and to write an edited caption that fixes the above mentioned problems (e.g. grammatical errors, awkward sentence structures, etc) in the initial one. If there are no errors in the initial caption, the annotator is instructed to only rephrase the initial caption. With this way we acquire a significant amount of linguistic corrections on the obtained captions and, in the same time, gather a new set of $N_{\text{c}}$ edited captions per audio file that offer variations in the structure of sentences and association of words. By crowdsourcing the task of the correction and rephrasing, we gain access to significantly large amount of workers (i.e. the users of the platform, compared to a non-crowdsourced solution) and diversity (because of the all different workers editing others' captions). It must be noted that if an annotator in the second step has provided a caption in the first step, then this annotator is not presented with his own caption(s) for editing.

After the first two steps each audio file has $N_{\text{c}}$ initial and $N_{\text{c}}$ edited captions, i.e. a total of $2N_{\text{c}}$ captions. These captions include initial captions with or without grammatical errors and edited captions that fail or succeed to remove errors. To determine which captions describe the audio most accurately and are grammatically most correct, we introduce the third step. In this step an annotator is presented with an audio file that is randomly selected from the audio files that have not yet been annotated with scores and each of the $2N_{\text{c}}$ captions of that audio file, and is asked to score each caption separately. The annotator scores each caption based on how accurately the caption describes the audio file (i.e. gives an accuracy score to the caption) and how grammatically correct it is (i.e. gives a fluency score to the caption). The annotator gives both scores on a scale of 1 to 4, with 1 meaning ``Bad'' and 4 meaning ``Very good''.
All the $2N_{\text{c}}$ captions of each audio file are scored by $N_{\text{t}}$ annotators. As in the previous step, annotators are not presented with their own captions to score. The total accuracy and fluency scores are then calculated by summing the $N_{\text{t}}$ individual scores. Finally, the $2N_{\text{c}}$ captions for each audio file are sorted in a descending fashion, accounting firstly for the total accuracy and then for the total fluency score. The top $N_{\text{c}}$ captions are selected for each audio file. 
%
%
\section{Evaluation}\label{sec:evaluation}
We evaluate our framework in the process of creating a new audio captioning dataset that will be released in Autumn 2019, by objectively assessing the impact of the three steps in terms of grammatical correctness and diversity of the gathered captions. We assess the grammatical correctness through the amount of typographical errors (the less errors, the better) and the diversity by examining the similarity of the captions (the less similar the captions, the more diversity). We use $N_{\text{a}}=5000$ audio files with time duration ranging from 15 to 30 seconds, and gathered randomly using the Freesound\footnote{\url{https://freesound.org/}} platform. Audio files in the Freesound platform are tagged, with tags indicating possible categories for the contents of each file. All $N_{\text{a}}$ files do not have tags indicating speech, music, and/or sound effects (e.g. ``bass'', ``glitch'', ``sci-fi''). All gathered audio files were post-processed to have a maximum absolute value of 1.
For each audio file we gather $N_{\text{c}}=5$ captions and at the third step, we employ $N_{\text{t}}=3$ annotators to rate the $2N_{\text{c}}$ captions, leading to a range of scores from 3 ($N_{\text{t}}* 1$) to 12 ($N_{\text{t}}* 4$). All annotations are gathered using the Amazon Mechanical Turk platform and its registered users as potential annotators. The first, second, and third steps are annotated by 693, 1033, and 1215 annotators with an average of 36, 24, and 12 annotations per annotator respectively. We present the obtained results in Section~\ref{sec:results}. 

We count the typographical errors appearing in the captions for each audio file separately for each of the $N_{\text{c}}$ initial, edited, final, and non-selected (in the third step) captions. To determine typographical errors we use the US and UK libraries of the CyHunspell python library\footnote{\url{https://pypi.org/project/CyHunspell/}}, which uses the Hunspell spellchecker\footnote{\url{http://hunspell.github.io/}}. Having edited captions with less errors than the initial captions measures the impact of the second step. Additionally, having a set of final selected captions with less typographical errors than the ones which are not selected, indicates that in the third step the framework provides a set of final captions that are better (grammatically) than the rest. 

To assess the diversity, we use the Jaccard similarity, also known as intersection over union. The Jaccard similarity of two sentences $a$ and $b$ is defined as
\begin{equation}\label{eq:jaq}
    J(a,b)=\frac{|\mathbb{W}_{a}\cap \mathbb{W}_{b}|}{|\mathbb{W}_{a}\cup \mathbb{W}_{b}|}\text{,}
\end{equation}
\noindent
where $\mathbb{W}_{a}$ is the set of stemmed words (i.e. words reduced to their roots, e.g. ``cat\textbf{s}'' to ``cat'') in sentence $a$, $\mathbb{W}_{b}$ is the set of stemmed words in sentence $b$, and $0\leq J(a,b)\leq 1$. When $J(a,b) = 0$, then the sentences $a$ and $b$ have no common (stemmed) words and the sets $\mathbb{W}_{a}$ and $\mathbb{W}_{b}$ are disjoint. On the contrary, $J(a,b) = 1$ shows that $\mathbb{W}_{a}$ and $\mathbb{W}_{b}$ contain exactly the same stemmed words. For word stemming (i.e. for finding the roots of words) we use the snowball stemmer from the NLTK language toolkit~\cite{loper:nltk}. To measure the amount of rephrasing, we calculate $J(a, b)$ between the initial and edited captions, using as $a$ each of the initial captions and as $b$ the corresponding edited caption. A high $J(a,b)$ will reveal almost no rephrasing and a low one will reveal significant rephrasing. To measure the diversity of the final $N_{\text{c}}$ captions, we firstly calculate the mean $J(a, b)$ for each audio file, using as $a$ and $b$ all the pairs of the final $N_{\text{c}}$ captions. Then, we calculate the mean $J$ across all audio files. We name the mean of the mean Jaccard similarity across all audio files as cross-similarity.

Finally, we evaluate the impact of the proposed framework on the set of words of all captions, that is, the final dictionary or word corpus that will be formed by all captions. We denote the set of words appearing in the $N_{\text{c}}$ captions of an audio file by $\mathbb{S}_{a}$. We merge all $\mathbb{S}_{a}$ to the multiset (i.e. bag) $\mathbb{S}_{T}$. We count the number of appearances of each of the words in $\mathbb{S}_{T}$, focusing on the rare words, i.e. words that appear up to five times in $\mathbb{S}_{T}$. For example, if a word in $\mathbb{S}_{T}$ has a number of appearances equal to two, it means that this word appears in the captions of exactly two audio files. This measure is of importance for the final dataset, because an audio file should not be in both the training and another split (i.e. validation or testing). This means that rare words that appear once in $\mathbb{S}_{T}$ result in unknown words/tokens to one of the splits. Words that appear twice in $\mathbb{S}_{T}$ result in audio files that can be used in two, different splits. 


%
%
\section{Results \& discussion}\label{sec:results}
Figure~\ref{fig:t2_typos} illustrates the frequency of audio files with typographical errors in their captions, for both initial and edited captions. It can be seen that the edited captions are less likely to contain any typographical errors than the initial captions. This means that the second step has a positive impact on the grammatical correctness, managing to produce captions with less typographical errors. In total, the edited captions have about 45\% less typographical errors than the initial captions.
\begin{figure}[!t]
    \centering
    \includegraphics[width=.8\columnwidth,trim={.55cm .2cm 1.1cm 1.35cm},clip]{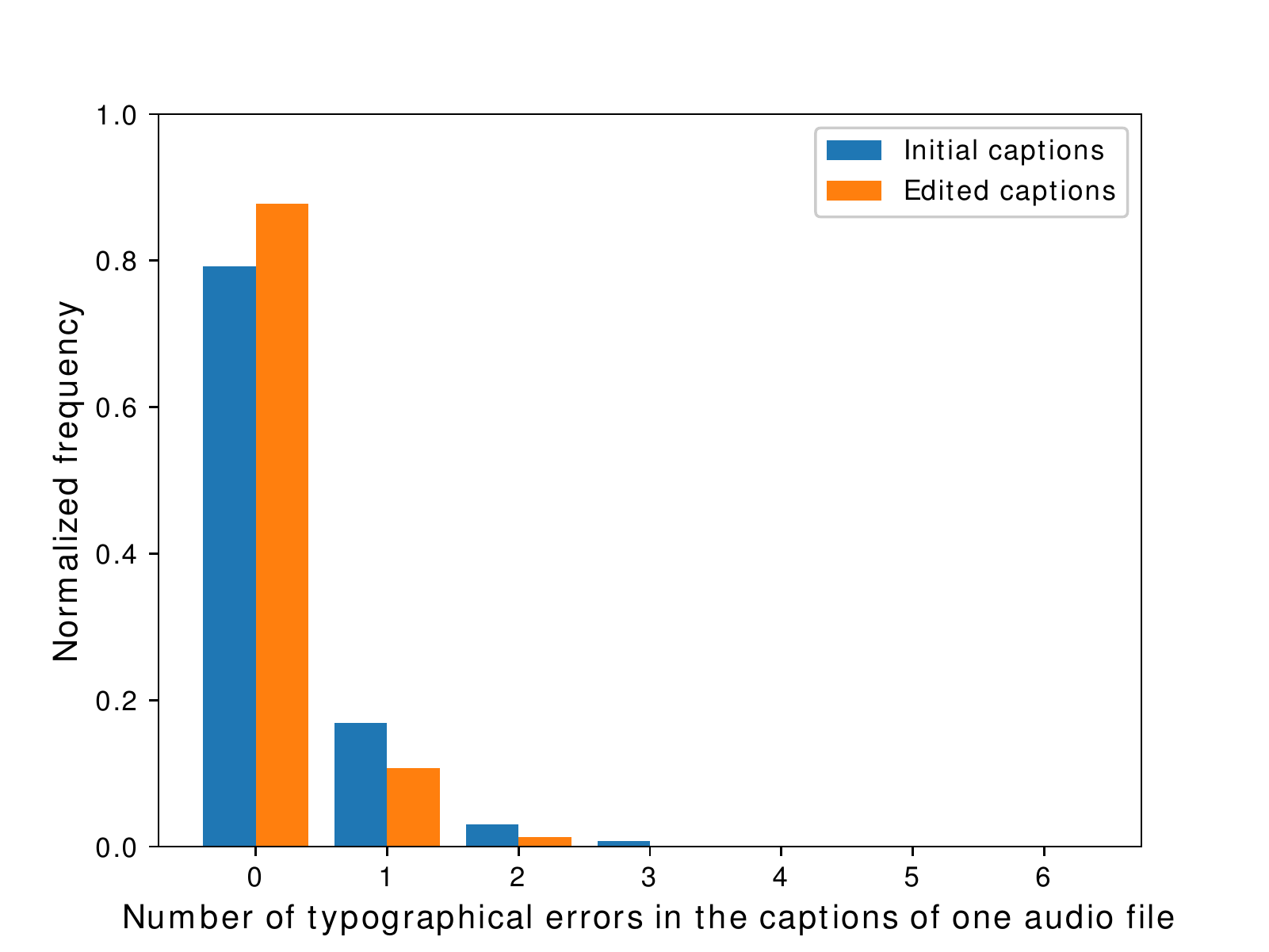}
    \caption{The number of typographical errors in the captions by the normalized frequency of audio files.}
    \label{fig:t2_typos}
\end{figure}
\begin{figure}[!t]
    \centering
    \includegraphics[width=.79\columnwidth,trim={0.2cm 0.2cm 1.5cm 1.34cm},clip]{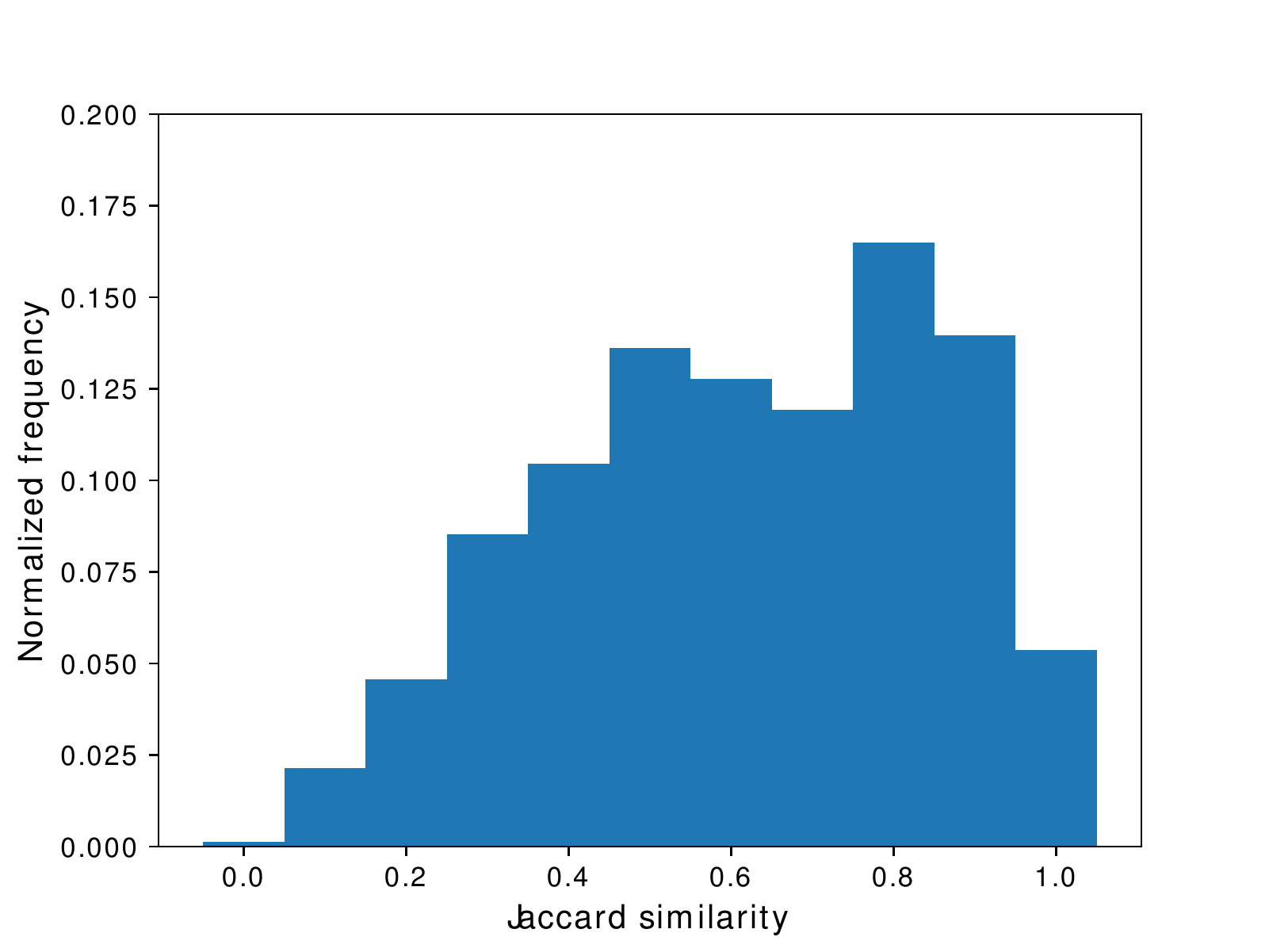}
    \caption{Jaccard similarity between initial and edited captions.}
    \label{fig:t2_jac_sim}
\end{figure}

Figure~\ref{fig:t2_jac_sim} illustrates the histogram of the Jaccard similarities between the initial and the corresponding edited captions. The average similarity is 0.62, which corresponds approximately to, e.g., changing two words in an 8-word caption. Therefore, the second step results in a reasonable amount of added diversity, roughly calculated to changing a fourth of the words in a sentence. Because the number of typographical errors in the edited captions is significantly lower than that of the initial captions, high frequencies of similarities in the high range might be a result of annotators fixing these errors and therefore not rephrasing the caption.

Figure~\ref{fig:t3_sim} displays a box plot of the cross-similarity values for the $N_{\text{c}}$ selected (i.e. the final), the other (i.e. the rest $N_{\text{c}}$ not selected at the third step), and the $N_{\text{c}}$ initial captions. The cross-similarity values for the selected, other, and initial captions are 0.24, 0.20, and 0.14 respectively. From the results it can be inferred that from the first step of our framework we indeed get a diverse set of initial captions. Moreover, the results in Figure~\ref{fig:t3_sim} show that the third step actually managed to control the increased diversity that the initial captions have, producing a lower (but still high) diversity for the captions. The baseline in the figure is calculated by creating random pairs of sentences, from all captions of all audio files.


\begin{figure}[!t]
    \centering
    \includegraphics[width=.79\columnwidth,trim={.5cm .5cm 1.5cm 1.2cm},clip]{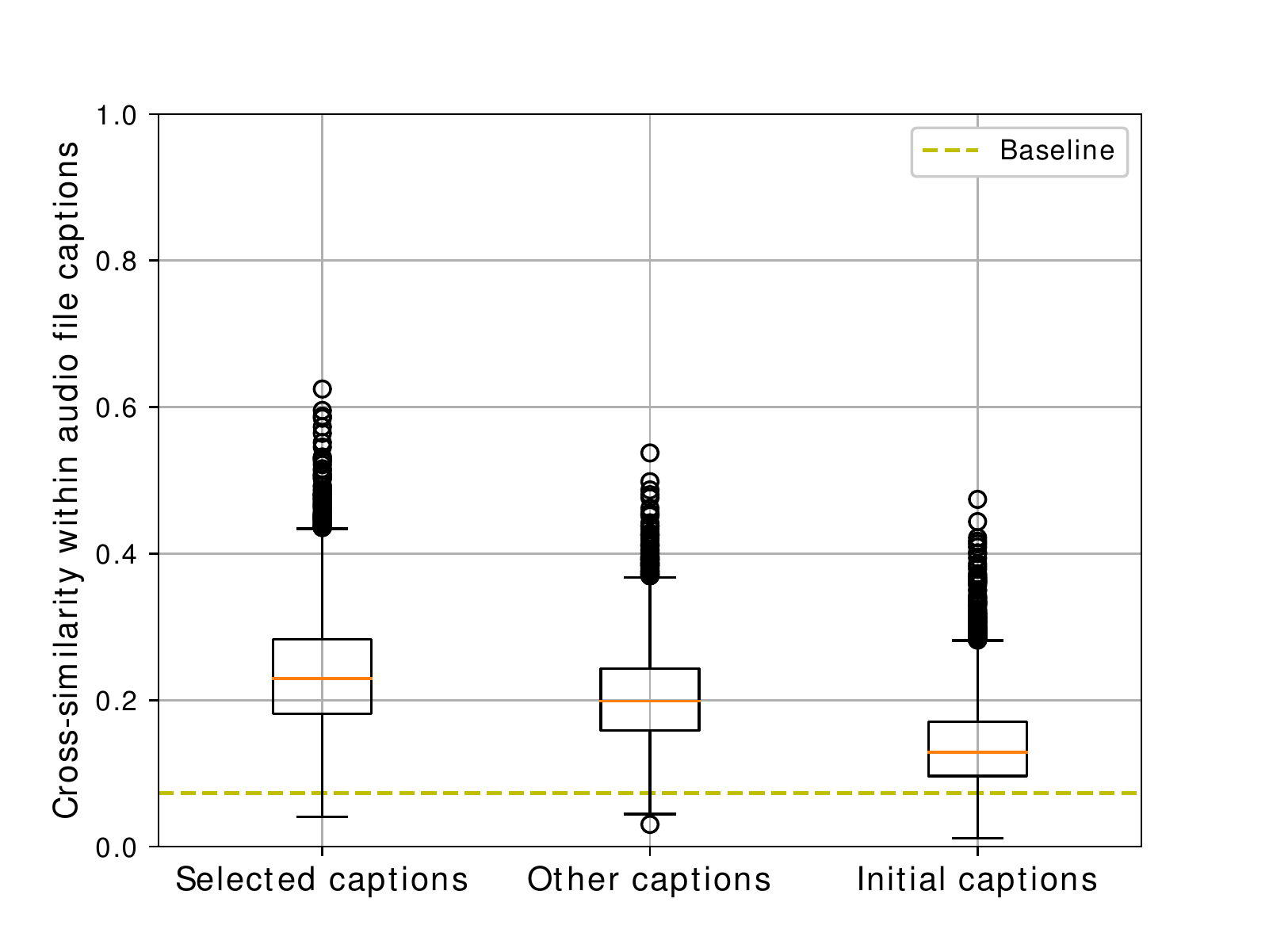}
    \caption{Cross-similarity of selected, non-selected, and initial captions}
    \label{fig:t3_sim}
\end{figure}
Figure~\ref{fig:t3_typos} depicts the percentage of captions versus the amount of typographical errors, considering also the total fluency score that is gathered in the third task for each, corresponding caption. It can be seen that the fluency score is inversely proportional to the amount of typographical errors. For example, the captions with a fluency score of 3 have, on average, 18 times more typographical errors than the captions with a fluency score of 12. These results clearly indicate that the fluency score successfully differentiated the levels of fluency within the captions.
\begin{figure}[!t]
    \centering
    \includegraphics[width=.82\columnwidth]{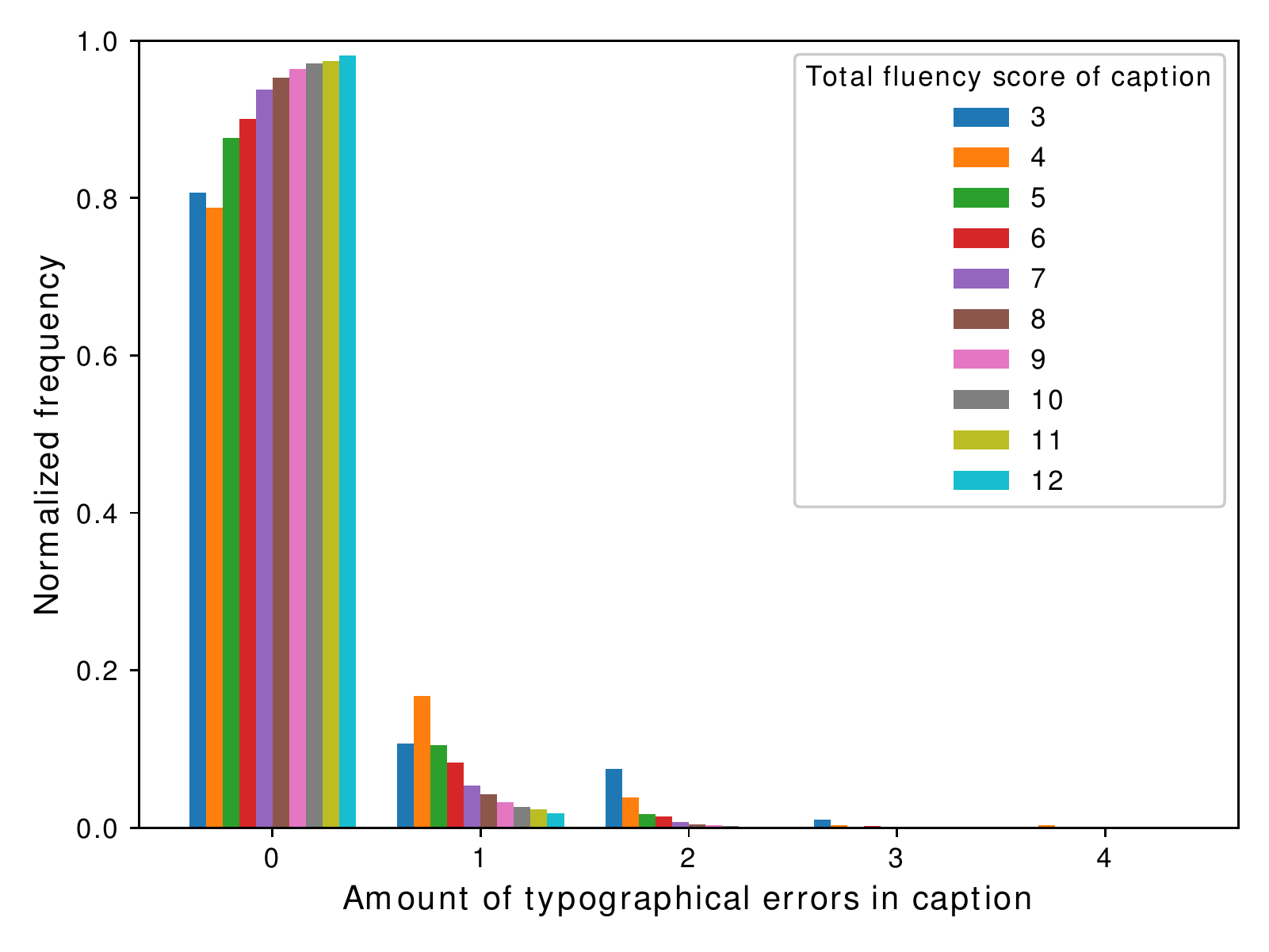} 
    \caption{The number of typographical errors appearing in a caption by the total fluency score given in the third step.}
    \label{fig:t3_typos}
\end{figure}
\begin{figure}[!t]
    \centering
    \includegraphics[width=.85\columnwidth]{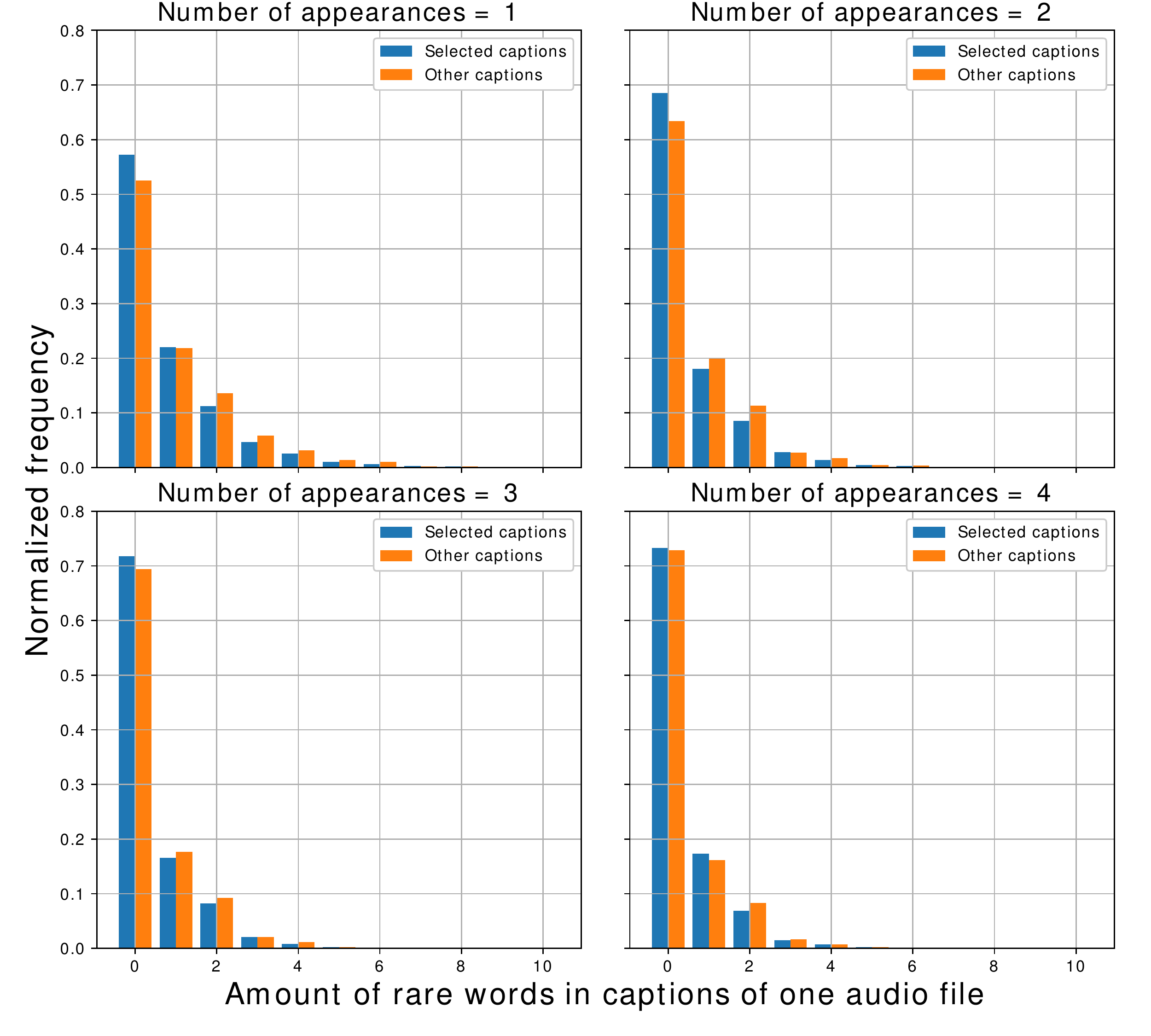} 
    \caption{The number of rare words with numbers of appearances of 1 to 4 by the normalized frequency of audio files.}
    \label{fig:t3_rare}
\end{figure}

Finally, in Figure~\ref{fig:t3_rare} are the percentages of audio files that have rare words, with numbers of appearances ranging from 1 to 4. The plots show that the selected captions are more likely not to contain any rare words with any number of appearances from one to four. This fact indicates that the resulting diversity imposed by the proposed framework does not hamper the quality of the resulting word corpus and the resulting dataset. 
%
%
\section{Conclusions \& future work}\label{sec:conclusions}
In this paper we presented a framework for the creation of an audio captioning dataset, using a crowdsourcing platform. Our framework is based on three steps of gathering, editing, and scoring the captions. We objectively evaluated the framework during the process of creating a new dataset for audio captioning, and in terms of grammatical correctness and diversity. The results show that the first step of our framework gathers a diverse set of initial captions, the second step gathers a set of edited captions that reduces the number of typographical errors in the initial captions while introducing additional diversity, and the third step extracts from the initial and edited captions a set of final selected captions that maintain a high diversity without introducing many rare words.

Further development of the framework could include pre-screening annotators as a way to eliminate manual screening of the annotations and automated processes for the control of more grammatical attributes and the amount of rare words.
%
\newpage
\clearpage
\bibliographystyle{IEEEtran}
\bibliography{refs}
\end{sloppy}
\end{document}